\documentclass{aa}

\usepackage[dvips]{graphicx}
\usepackage{txfonts}
\usepackage{times,epsfig}
\usepackage{natbib} 
\bibpunct{(}{)}{;}{a}{}{,}
 
\newcommand{\Msun}{\mathrm{M}_{\odot}}
\newcommand{\Zsun}{\mathrm{Z}_{\odot}}
\newcommand{\Lsun}{\mathrm{L}_{\odot}}

\begin{document}
%
%---------------------------------------------------------------------------------------------------------------------------------
% Title 
%---------------------------------------------------------------------------------------------------------------------------------
%
   \title{Genesis and evolution of dust in galaxies in the early Universe}
   
   \subtitle{ II. Rapid dust evolution in quasars at $z$ $\gtrsim$ 6}
%
%---------------------------------------------------------------------------------------------------------------------------------
% Authors & Institutions
%---------------------------------------------------------------------------------------------------------------------------------
%
   \author{C. Gall \inst{1}, A. C. Andersen \inst{1},
          \and
          J. Hjorth \inst{1}
          }

   \institute{\inst{1}Dark Cosmology Centre, University of Copenhagen, Niels Bohr Institute, Juliane Maries Vej 30, DK-2100 Copenhagen, Denmark
%                  \\
%              \email{christa@dark-cosmology.dk}\\
%               \email{anja@dark-cosmology.dk}\\
%               \email{jens@dark-cosmology.dk}
             }

   \date{Received January 07, 2011}
%   
%---------------------------------------------------------------------------------------------------------------------------------
% Abstract                  
%---------------------------------------------------------------------------------------------------------------------------------
%   
  \abstract
  % context heading (optional), leave it empty if necessary  
   {}
   {We intend to assess the most plausible scenarios for generating large amounts of dust in high-$z$ quasars (QSOs) 
   on the basis of observationally derived physical properties of QSOs at $z$ $\gtrsim$ 6.  
   }
   {We use a chemical evolution model to compute the temporal progression of quantities such as
   the amount of dust and gas, stellar masses, star formation rates (SFRs) and the metallicity
   for various combinations of the initial mass function (IMF), the mass of the galaxy, 
   dust production efficiencies, and the degree of dust destruction in the ISM. 
   We investigate the influence of the SFR on the evolution of these quantities, and 
   determine the earliest epochs at which agreement with observations can be achieved. 
  We apply the obtained results to individual  QSOs at $z$ $\gtrsim$ 6.
  }
   {We find that large quantities of dust can be generated rapidly 
   as early as 30 Myr after the onset of the starburst when the SFR of the starburst 
   is $\gtrsim$ $10^{3}$ $\Msun$ yr$^{-1}$. 
   The amount of dust and several other physical quantities of individual QSOs at $z$ $\gtrsim$ 6 
   are satisfactorily reproduced 
   by models at epochs 30, 70, 100, and 170 Myr for 
    galaxies with initial gas masses of 1--3 $\times$  $10^{11}$ $\Msun$. The best agreement with 
    observations is obtained with top-heavy IMFs.
    A sizable dust contribution from supernovae (SNe) is however required,  
    while at these epochs dust production by asymptotic giant branch (AGB) stars is negligible. 
    Moderate dust destruction in the ISM can be accommodated. 
     }
   {}

                     
  \titlerunning{Genesis and evolution of dust in galaxies in the early Universe}    
 \authorrunning{C.Gall et al}             
   \maketitle
%
%---------------------------------------------------------------------------------------------------------------------------------
% Section 1: Introduction 
%---------------------------------------------------------------------------------------------------------------------------------
%   
\section{Introduction}
%---------------------------------------------------------------------------------------------------------------------------------
\defcitealias{gall10a}{Paper I}
%---------------------------------------------------------------------------------------------------------------------------------
%
Studying QSOs and their host galaxies at high redshift ($z$ $>$ 6) is important to gain deeper 
insight into the formation and evolution of galaxies, the origin of dust production, and the build 
up of stellar bulge masses in coevolution with  supermassive black holes (SMBHs). 

While the most distant known QSO, J114816.64+525150.3  
\citep[][herafter J1148+5251]{fan03}, is at $z$ = 6.4, several tens of QSOs have been discovered at 
$z$ $\sim$ 6  \citep[e.g.,][]{fan04, fan06, will07, jiang10}. 
Most of the observed QSOs at this redshift, where the epoch of cosmic evolution is $\sim$ 1 Gyr, 
exhibit extreme physical properties such as very high far-infrared (FIR) luminosities which imply  
large dust masses  \citep[e.g.,][]{omo01, omo03, caril01, bertol02}, 
and SMBHs with masses $>$ 10$^{9}$ $\Msun$  \citep[e.g.,][]{bar03, will03, vest04}. 

Observations of QSOs have shown that dust emission at near-infrared (NIR) wavelengths arise 
from warm and hot dust ($T$ $\lesssim$ 1000 K) assembled within a few parsec  \citep[e.g.,][]{hin06,jiang06}. 
The NIR emission is believed to be powered by the active galactic nucleus and related to the 
QSO activity \citep[e.g.,][]{poll00}.  
However, two QSOs at $z$ $\sim$ 6 without detectable emission from hot dust have been found 
 \citep{jiang06, jiang10}.  
It has been proposed that these QSOs are at a too early evolutionary stage to have built up 
significant amounts of hot dust. 
Alternative scenarios including for example the destruction of the hot dust or 
dust misalignments from the SMBH have also been discussed \citep{hao10a, hao10b, gued10}.

The FIR luminosity of $L_{\mathrm{FIR}}$ $\sim$ 10$^{12-13}$ $\Lsun$ is attributed to cold dust 
($T$ $\sim$ 30--60 K)  \citep[e.g.,][]{wan08} which is probably distributed over kilo-parsec scales 
throughout the host galaxy \citep{leip10}. The amount of cold dust inferred is about a few times 
10$^{8}$ $\Msun$  \citep[e.g.,][]{bertol03, robs04, beel06, mich10b}. 
The dominant  source of the high FIR luminosity is believed to be dust heated by intense star 
formation in the circumnuclear region  \citep[e.g.,][]{caril04, rie07, wan08}. 
Detection of [\ion{C}{ii}] line emission at 158 $\mu$m  \citep{mai05} within a central region with 
radius $\sim$ 750 pc of the host galaxy of J1148+5251 also implies a high star formation rate 
surface density of 1000 $\Msun$ yr$^{-1}$ kpc$^{-2}$ \citep{walt09}. 
\citet{wan10} derived SFRs between 530--2300 $\Msun$ yr$^{-1}$ from observations of a sample 
of QSOs at redshift $z$ $>$ 5.   
Observations of strong metal emission of high-$z$ QSOs  \citep[e.g.,][]{bar03, diet03, mai03, bec06} 
indicate strong star forming activity in the QSO hosts and solar or supersolar metallicity
 \citep[e.g.,][]{fan03, freu03, juar09}.  
Theoretical studies of the gas metallicity of QSO hosts 
also predict supersolar metallicities for $z$ = 5--6 QSOs \citep[e.g.,][]{dimat04}. 

The high inferred SFRs imply short timescales ($\le 10^{8}$ yr) of the starburst 
\citep[e.g.][]{bertol03, walt04, dwe07, rie09}, and consequently a young age of the QSOs. 
An early evolutionary stage of  $z$ $>$ 4 QSOs has also been suggested from studies of extinction 
curves of broad absorption line QSOs \citep[e.g.,][]{galler10} 
which turned out to be best fitted with extinction curves for SN-like dust  \citep[e.g.,][]{mai04, mai06, galler10}. 
This suggests SNe as the preferential source of dust at early epochs  \citep[e.g.,][]{dwe98, mor03, hiras05, dwe07, dwe10}, 
even though the dust productivity of SNe is poorly constrained  (for a review see Gall et al. in prep).   
The dust in high-$z$ QSOs could also be grown in the ISM \citep[e.g.,][]{drai09, mich10b, pip10}.
Finally, a dominant dust production by asymptotic giant branch stars has been claimed  \citep[][]{val09}.

Molecular gas masses of the order of $\sim$ 1--2.5 $\times$ $10^{10}$ $\Msun$ have been inferred 
from detections of high excitation CO line emission in QSOs at $z$ $>$ 5 within a  $\sim$ 2.5 kpc radius 
region  \citep[e.g.,][]{bertol03b, walt03, walt04, wan10}. 
The dynamical masses inferred from these CO observations are a few times 
$\sim$ $10^{10-11}$ $\Msun$ which sets an upper limit on stellar bulge masses. 
These however are roughly two orders of magnitude lower than required from the present 
day black hole-bulge relation \citep[e.g.,][]{mah03}. 
It therefore has been proposed that the formation of the SMBH occurs prior to the formation of 
the stellar bulge. QSOs will then have to accrete additional material to build up the required bulge mass 
 \citep[e.g.,][]{walt04, rie09, wan10}.  
For QSOs at $z$ $>$ 6 super-Eddington growth on timescales shorter than $\sim$10$^{8}$ yr 
seem to be required to form a SMBH $>$ 10$^{9}$ $\Msun$  \citep[e.g.,][]{kaw09}.
It has also been predicted that QSOs at $z$ $\sim$ 6 likely have formed in dark matter halos of 
10$^{12-13}$ $\Msun$ \citep[e.g.,][]{li07, kaw09}.

In \citet[][herafter \citetalias{gall10a}]{gall10a} we developed a chemical evolution model to elucidate the conditions required 
for generating large dust masses in high-$z$ starburst galaxies. 
We showed that galaxies with masses of 
1--5  $\times$ 10$^{11}$ $\Msun$ are suitable for enabling the production of large amounts of dust within $\sim$ 400 Myr.
In the present  paper we apply this model to QSOs at $z$ $\gtrsim$ 6. 
We perform more detailed comparison between model results and values inferred from observations 
of $z$ $\gtrsim$ 6 QSOs to identify the most likely scenario. 
Furthermore, we consult additional parameters such as the H$_2$ mass and the CO conversion factor for more refined evaluations. 
In particular, calculations with higher SFRs than in \citetalias{gall10a} are considered. 
We aim to determine the earliest epochs at which the model results are in agreement with 
those from observations.  
  
The structure of the paper is as follows: In Sec. \ref{SEC:MOD} we briefly review the model developed 
in \citetalias{gall10a}.  
A detailed analysis of the results is presented in Sec. \ref{SEC:LOSB} followed by a discussion in 
Sec. \ref{SEC:DISC}.    
%
 %
%---------------------------------------------------------------------------------------------------------------------------------
% Section : The model 
%---------------------------------------------------------------------------------------------------------------------------------
%   
\section{The model}
\label{SEC:MOD}
%---------------------------------------------------------------------------------------------------------------------------------
%
The galactic chemical evolution model from  \citetalias{gall10a} 
 is self-consistent,  numerically solved and has been developed to ascertain the temporal 
progression of dust, gas, metals, and diverse physical properties of starburst galaxies. 
The incorporated stellar sources are AGB stars in the mass range 3--8 $\Msun$ and SNe. 
A differentiation between diverse SN subtypes has been implemented.   
Their roles as sources of dust production, dust destruction or suppliers of gas and 
heavy elements are taken into account.
The lifetime dependent yield injection by the stellar sources, as well as dust destruction 
in the ISM due to SN shocks are also taken into account.
Moreover, the formation of a SMBH is considered.
Due to the very high SFRs of the starbursts, infall of neutral gas will only effect the system for comparable high infall rates. 
Thus, gas infall and outflows are neglected. Possible caveats of such an approach are discussed in  \citetalias{gall10a}.
The model allows investigations of a broad range of physical properties of galaxies.

The prime parameters are summarized in the following.
\begin{itemize}
\item Three different possible prescriptions for the stellar yields of SNe are implemented, 
i.e., (i) stellar evolution models by \citet{eld08} 
(referred to as `EIT08M'), (ii) rotating stellar models by \citet{geo09}, 
or (iii) nucleosynthesis models by either \citet{woos95} or \citet{nom06}. 
The stellar yields for AGB stars are taken from \citet{vhoek97}.     
 
 \item We differentiate between five different IMFs. 
These are a \citet{salp55} IMF, a top-heavy, and a mass-heavy IMF, as well as 
IMFs  \citep{lars98} with characteristic masses of either 
$m_{\mathrm{ch}}$ = 0.35 (Larson 1) or $m_{\mathrm{ch}}$ = 10 (Larson 2). 

\item The SFR at a certain epoch is given by the Kennicutt law  \citep{kenn98} as 
          $\psi(t) = \psi_{\mathrm{ini}} \, (M_{\mathrm{ISM}}(t) / M_{\mathrm{ini}})^k$, where 
          $\psi_{\mathrm{ini}}$ is the initial SFR of the starburst, $M_{\mathrm{ISM}}(t)$ is the 
          initial gas mass of the galaxy and $k$ = 1.5. 

\item The amount of dust produced by SNe and AGB stars is calculated using the 
dust formation efficiencies discussed in \citetalias{gall10a}. 
For SNe three different dust production efficiency limits are determined, i.e. a `maximum' SN efficiency, 
a `high' SN efficiency,  and a `low' SN efficiency. 
The `maximum' SN efficiency originates from theoretical SN dust formation models, 
and corresponds to dust masses of approximately 3--10 $\times$ 10$^{-1}$ $\Msun$. 
Similar dust masses have been observed in SN remnants such as Cas A \citep[e.g.,][]{dun09} or Kepler \citep[e.g.,][]{gom09}. 
Dust destruction in reverse shock interaction of about 93 \% has been applied to the `maximum' 
SN efficiency, to obtain the `high' SN efficiency. 
The amount of dust for instance is $\sim$  2--6 $\times$ 10$^{-2}$ $\Msun$, 
which is also comparable to some observations of older SN remnants \citepalias[][see references therein]{gall10a}.  
The `low' SN efficiency is based on SN dust yields (on average about 3 $\times$ 10$^{-3}$ $\Msun$) inferred from observations of SN ejecta. 

\item Dust destruction in the ISM is implemented in terms of the mass of ISM material, 
$M_{\mathrm{cl}}$, swept up by a single SN shock and cleared of the containing dust.
\end{itemize}

For calculations in this paper  most parameters have the same settings as defined in 
 \citetalias[][]{gall10a}. 
We apply the models where the formation of a SMBH has been included. 
A constant growth rate has been estimated based on the final mass of the SMBH and the considered growth timescale. 
In this paper  the SMBH growth is considered with a shorter growth timescale and calculations are performed with higher initial SFRs. 
For the SN yields we only consider the case of EIT08M.
The parameters which differ from those used in \citetalias{gall10a} are listed in 
Table~\ref{TAB:PAR}.

\begin{table}%[!t]
\caption{{Model parameters}}            
\label{TAB:PAR}
\centering
\begin{tabular}{llll}
\hline
\hline
   Parameters			& Value									& Unit				&Description\\                 
\hline
\\   $\psi_{\mathrm{ini}}$    & 3 $\times$ $10^3$, 1 $\times$ $10^4$			& $\Msun$  yr$^{-1}$	& Star formation rate\\  
   $M_{\mathrm{SMBH}}$	& 3 $\times$  $10^{9}$						&  $\Msun$			& Mass of the SMBH\\
  $t_{\mathrm{SMBH}}$  	& 1 $\times$ $10^{8}$						& yr					& Growth timscale\\ 
  					&										&					& for the SMBH\\
\hline
\end{tabular}\\
\end{table}
%\ \\
%
 %---------------------------------------------------------------------------------------------------------------------------------
% Section : Late onset of starburst / Results
%---------------------------------------------------------------------------------------------------------------------------------
%
\section{Results}
\label{SEC:LOSB}
%---------------------------------------------------------------------------------------------------------------------------------
%
%
    \begin{figure*}
      \centering
   \includegraphics[width=12cm]{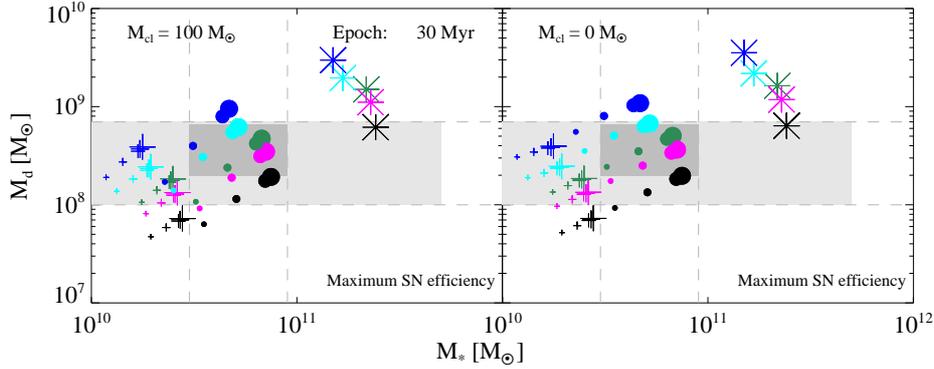}   
      \caption{ Relation between dust mass and stellar mass at an epoch of 30 Myr, 
      for various initial gas masses and IMFs. 
      Calculations are performed for a `maximum' SN efficiency and dust destruction 
      in the ISM with $M_{\mathrm{cl}}$ = 100 $\Msun$ (left panel) and 
      $M_{\mathrm{cl}}$ = 0 (right panel).
      The colored symbols are obtained for different initial gas masses, $M_{\mathrm{ini}}$, 
      SFRs, and IMFs.
      The size of the symbols is scaled by $M_{\mathrm{ini}}$. Calculations are made for   
       $M_{\mathrm{ini}}$ = 1.3 $\times$  $10^{12}$ $\Msun$ (largest symbol), 
       $M_{\mathrm{ini}}$ = 5 $\times$  $10^{11}$ $\Msun$,
       $M_{\mathrm{ini}}$ = 3 $\times$  $10^{11}$ $\Msun$,
       $M_{\mathrm{ini}}$ = 1 $\times$  $10^{11}$ $\Msun$, and
       $M_{\mathrm{ini}}$ = 5 $\times$  $10^{10}$ $\Msun$ (smallest symbol). 
      The crosses correspond to calculations 
      for a initial SFR  $\psi_{\mathrm{ini}}$ =  $10^{3}$ $\Msun$ yr$^{-1}$, 
      the filled circles to $\psi_{\mathrm{ini}}$ = 3 $\times$ $10^{3}$ $\Msun$ yr$^{-1}$, 
      and the stars to $\psi_{\mathrm{ini}}$ = $10^{4}$ $\Msun$ yr$^{-1}$. 
      The black, green, cyan, magenta, and blue colors denote the Salpeter, mass-heavy, top-heavy, 
      Larson 1, and Larson 2 IMF, respectively. 
      The dark grey region indicates the mass range of  stellar masses and  dust masses
      derived from observations of QSOs at $z$ $>$ 6. The vertical dashed lines represent the 
      lower and upper limits of the observed stellar masses.
      The light grey area illustrates the whole mass ranges derived from observations of 
      QSOs $>$ 5 and accounts for uncertainties in the derived quantities.
      The horizontal dashed lines mark the lower and upper mass limit of the derived dust masses.  
                      }
     \label{FIG:DUSE30}
   \end{figure*}
    \begin{figure*}
   \centering
   \includegraphics[width=\textwidth]{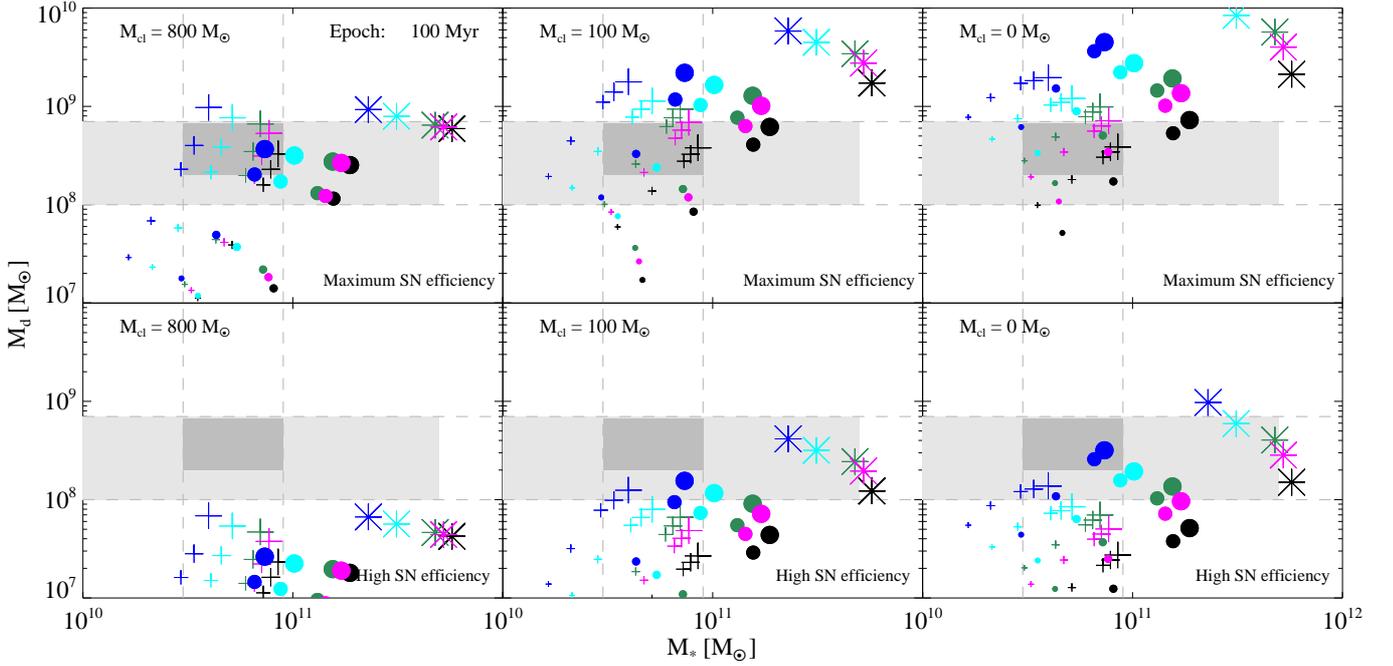}   
      \caption{Relation between dust mass and stellar mass at an epoch of 100 Myr for 
      various initial gas masses and IMFs. 
      Calculations are performed for a `maximum' SN efficiency (top row)  
      and a `high' SN efficiency (bottom row). 
      Dust destruction in the ISM is considered for a  
      $M_{\mathrm{cl}}$ = 800 $\Msun$ (left column), 
      $M_{\mathrm{cl}}$ = 100 $\Msun$ (middle column), 
      and $M_{\mathrm{cl}}$ = 0 (right column).
      The colored symbols are
      obtained for different initial gas masses, $M_{\mathrm{ini}}$, SFRs, and IMFs.
      The size of the symbols is scaled by $M_{\mathrm{ini}}$.  Calculations are made for   
       $M_{\mathrm{ini}}$ = 1.3 $\times$  $10^{12}$ $\Msun$ (largest symbol), 
       $M_{\mathrm{ini}}$ = 5 $\times$  $10^{11}$ $\Msun$,
       $M_{\mathrm{ini}}$ = 3 $\times$  $10^{11}$ $\Msun$,
       $M_{\mathrm{ini}}$ = 1 $\times$  $10^{11}$ $\Msun$ and
       $M_{\mathrm{ini}}$ = 5 $\times$  $10^{10}$ $\Msun$ (smallest symbol). 
      The crosses correspond to calculations for a initial SFR  
      $\psi_{\mathrm{ini}}$ =  $10^{3}$ $\Msun$ yr$^{-1}$, the filled circles to 
      $\psi_{\mathrm{ini}}$ = 3 $\times$ $10^{3}$ $\Msun$ yr$^{-1}$, 
      and the stars to $\psi_{\mathrm{ini}}$ = $10^{4}$ $\Msun$ yr$^{-1}$. 
      The black, green, cyan, magenta, and blue colors denote the Salpeter, 
      mass-heavy, top-heavy, Larson 1, and Larson 2 IMF, respectively. 
      The dark grey region indicates the mass range of stellar masses and dust masses 
      derived from observations of QSOs at $z$ $>$ 6. 
      The vertical dashed lines represent the lower and upper limits of the observed stellar masses.
      The light grey area illustrates the whole mass ranges derived from observations of 
      QSOs $>$ 5 and accounts for uncertainties in the derived quantities.
      The horizontal dashed lines mark the lower and upper mass limit of the derived dust masses.                        
      }
     \label{FIG:DUSE}
   \end{figure*}
In this section we present the results of models calculated within short timescales
after the starburst. 

A short enrichment timescale of a few times 10$^7$ yr for an intense starburst with a 
SFR of $\sim$ 3 $\times$ $10^{3}$ $\Msun$ yr$^{-1}$ has been proposed 
by  e.g., \citet{bertol03}, \citet{walt04}, \citet{dwe07}, \citet{rie09}. 
Owing to this suggestion we are interested in whether the observed large 
dust masses  in excess of $10^{8}$ $\Msun$ can be reached within about 100 Myr. 
Consequently we performed calculations with an initial SFR for the starburst with 
$\psi_{\mathrm{ini}}$ = 3 $\times$ $10^{3}$ $\Msun$ yr$^{-1}$ 
for galaxies with initial gas masses 
$M_{\mathrm{ini}}$ = 5 $\times$  $10^{10}$ $\Msun$, 
$M_{\mathrm{ini}}$ = 1 $\times$  $10^{11}$ $\Msun$, 
$M_{\mathrm{ini}}$ = 3 $\times$  $10^{11}$ $\Msun$, and 
$M_{\mathrm{ini}}$ = 5 $\times$  $10^{11}$ $\Msun$. 
For the most massive system with 
$M_{\mathrm{ini}}$ = 1.3 $\times$  $10^{12}$ $\Msun$ 
an initial SFR 
$\psi_{\mathrm{ini}}$ = 1 $\times$ $10^{4}$ $\Msun$ yr$^{-1}$ is adopted.  
We included the results for a lower initial SFR of 
$10^{3}$ $\Msun$ yr$^{-1}$ from 
models computed in \citetalias{gall10a} for comparison.  

In \citetalias{gall10a} we analyzed the evolution of the amount of dust and 
various physical properties, 
and found that these are strongly dependent on the mass of the galaxy. 
Moreover, for a given initial SFR all quantities evolve faster in less massive 
galaxies. 
In this paper we perform detailed comparisons between calculated and observed values of 
the total dust mass, $M_{\mathrm{d}}$, the stellar mass, 
$M_{\mathrm{\ast}}$, the SFR, $\psi$, and the metallicity, $Z$. 
We identified the shortest epoch, where some model results are in accordance with 
observations to be 30 Myr. 
Furthermore, we discuss quantities such as the CO conversion factor, the gas-to-H$_{2}$ mass ratio, 
and the possible amount of molecular hydrogen. 
\subsection{Dust and stellar mass}
%---------------------------------------------------------------------------------------------------------------------------------
%
%
In Fig.~\ref{FIG:DUSE30} we present the results for the mass of dust  
versus the stellar mass  for galaxies with different initial gas masses 
and initial SFRs at an epoch of 30 Myr.
The displayed models are computed for a `maximum' SN efficiency. 
Dust destruction in the ISM is considered for values of 
$M_{\mathrm{cl}}$ = 100 $\Msun$ (left panel) and $M_{\mathrm{cl}}$ = 0 (right panel). 

The dark grey region represents the mass ranges of the stellar mass and dust mass 
derived from observations of QSOs at $z$ $>$ 6.
The lower and upper limits of the stellar mass are estimated by subtracting the molecular 
gas masses, $M_{\mathrm{H_{2}}}$ from the total dynamical masses, $M_{\mathrm{dyn}}$.
Values for $M_{\mathrm{dyn}}$ and $M_{\mathrm{H_{2}}}$ are based on data from  
\citet[][and references therein]{wan10} for three QSOs at $z$ $>$ 6. 
For an estimation of $M_{\mathrm{dyn}}$ an inclination angle $i = 65\degr$ of the gas disk 
is taken for QSO J1148+5251 \citep{walt04}, while $i = 40\degr$ similar to \citet{wan10} is 
applied to the remaining two QSOs. 
We adopt the lower and upper limits for the dust masses  from \citet{beel06} and \citet{mich10b}.
The light grey region covers the range of derived stellar masses and dust masses from 
observations of QSOs $>$ 5 \citep{wan10, mich10b}. 
The boundaries for the stellar masses are estimated similar to the QSOs at $z$ $>$ 6 
(with $i = 40\degr$ for deriving $M_{\mathrm{dyn}} $).
We set the lower dust limit to $10^{8}$ $\Msun$ to account for the uncertainties of 
derived dust masses from observations.

Despite the short time span of 30 Myr, it is evident that most models are within the 
plausible mass ranges illustrated by the light and dark grey regions. 
This signifies a rapid build-up of a large amount of dust, provided SNe produced dust with
a `maximum' SN efficiency. 
For galaxies with $M_{\mathrm{ini}}$ = 1--5 $\times$  $10^{11}$ $\Msun$ 
all models with an initial SFR of 3 $\times$ $10^{3}$ $\Msun$ yr$^{-1}$ are in agreement 
with the observed values for the stellar masses 
for QSOs at $z$ $>$ 6. 
The requirements for $M_{\mathrm{d}}$ are best accomplished with either a top-heavy, 
mass-heavy or Larson 1 IMF for both values of  $M_{\mathrm{cl}}$. 
In a galaxy with $M_{\mathrm{ini}}$ = 1 $\times$  $10^{11}$ $\Msun$ the amount of dust reached 
with a Larson 2 IMF  and $M_{\mathrm{cl}}$ = 100 $\Msun$ also matches with the dark grey region. 
Models for either a `high' or `low' SN efficiency did not reach $10^{8}$ $\Msun$ of dust. 
Only in the most massive galaxy ($M_{\mathrm{ini}}$ = 1.3 $\times$  $10^{12}$ $\Msun$) 
and for top-heavy IMFs with a `high' SN efficiency an amount of dust $>$ $10^{8}$ $\Msun$ is obtained.

In Fig.~\ref{FIG:DUSE} we illustrate the results for dust and stellar masses at an epoch of 100 Myr. 
We present models for a `maximum' SN efficiency (top row) and a `high' SN efficiency (bottom row), 
while dust destruction in the ISM is considered for a $M_{\mathrm{cl}}$ = 800 $\Msun$ (left column), 
$M_{\mathrm{cl}}$ = 100 $\Msun$ (middle column), and $M_{\mathrm{cl}}$ = 0 (right column). 
We carried out calculations for a `low' SN efficiency, but the obtained dust masses of these 
models remained below  $10^{8}$ $\Msun$. 

At these early epochs the stellar mass, $M_{\mathrm{\ast}}$, is higher for models with an initially larger SFR (at fixed IMF and $M_{\mathrm{ini}}$). The stellar mass is also larger for IMFs biased towards low mass stars (at fixed $M_{\mathrm{ini}}$ and $\psi_{\mathrm{ini}}$).
It is interesting to note that in the less massive galaxies (0.5--1 $\times$  $10^{11}$ $\Msun$) 
dust masses obtained for the higher initial SFR 
($\psi_{\mathrm{ini}}$ = 3 $\times$ $10^{3}$ $\Msun$ yr$^{-1}$) are lower than dust masses 
obtained for the lower SFR ($\psi_{\mathrm{ini}}$ = $10^{3}$ $\Msun$ yr$^{-1}$). 
Moreover, in these galaxies the amount of dust reached at an epoch of 30 Myr 
(see Fig.~\ref{FIG:DUSE30}) and  for  $M_{\mathrm{cl}}$ = 100--800 $\Msun$ is also higher 
than that seen at the epoch of 100 Myr for same $M_{\mathrm{cl}}$. 

We find that the stellar masses for models with an initial SFR 
$\psi_{\mathrm{ini}}$ = 1--3 $\times$ $10^{3}$ $\Msun$ yr$^{-1}$ 
are within the observed region for $z$ $>$ 5 QSOs.
For some models with 
$\psi_{\mathrm{ini}}$ = 3 $\times$ $10^{3}$ $\Msun$ yr$^{-1}$, 
stellar masses are within the mass range for $z$ $>$ 6 QSOs. 
This in particular applies to systems with either $M_{\mathrm{ini}}$ = 0.5--1 $\times$  $10^{11}$ $\Msun$ 
(all IMFs) or $M_{\mathrm{ini}}$ = 3--5 $\times$  $10^{11}$ $\Msun$ with top heavy IMFs. 
Stellar masses within the dark grey area are also found with 
$\psi_{\mathrm{ini}}$ = $10^{3}$ $\Msun$ yr$^{-1}$ for galaxies with either 
$M_{\mathrm{ini}}$ = 3--13 $\times$  $10^{11}$ $\Msun$ and top heavy IMFs or for the 
less massive galaxies in combination with IMFs favoring low mass stars. 

In the case of $M_{\mathrm{cl}}$ = 800 $\Msun$ and for a `maximum' SN efficiency most 
models with $M_{\mathrm{ini}}$ = 3--13 $\times$  $10^{11}$ $\Msun$ and 
$\psi_{\mathrm{ini}}$ = $10^{3}$ $\Msun$ yr$^{-1}$  fit within the dark grey region. 
However for the higher initial SFR $M_{\mathrm{d}}$ is within or close to this zone only for 
galaxies with $M_{\mathrm{ini}}$ = 3--5 $\times$  $10^{11}$ $\Msun$ and top-heavy IMFs. 
For $M_{\mathrm{cl}}$ = 100 $\Msun$ and a `maximum' SN efficiency the dust mass obtained 
in a galaxy  with $M_{\mathrm{ini}}$ = 1 $\times$  $10^{11}$ $\Msun$, 
$\psi_{\mathrm{ini}}$ = 3 $\times$ $10^{3}$ $\Msun$ yr$^{-1}$ and for top-heavy IMFs 
is in agreement with observations, while the dust masses in the more massive galaxies 
for some IMFs and SFRs are higher than required. 
In the case of no dust destruction the dust masses reached 
for some IMFs and SFRs are able to match within the dark grey area also in the least massive galaxy.  

We find that in case of a `high' SN efficiency and for 
$\psi_{\mathrm{ini}}$ = 3 $\times$ $10^{3}$ $\Msun$ yr$^{-1}$ in galaxies with initial masses  
3--5 $\times$ $10^{11}$ $\Msun$ and top-heavy IMFs high dust masses are possible, 
even if dust destruction is included (i.e., $M_{\mathrm{cl}}$ = 0--100 $\Msun$).  
\subsection{Metallicity and SFR}
%---------------------------------------------------------------------------------------------------------------------------------
%
    \begin{figure}
   \centering
  \resizebox{\hsize}{!}{ \includegraphics{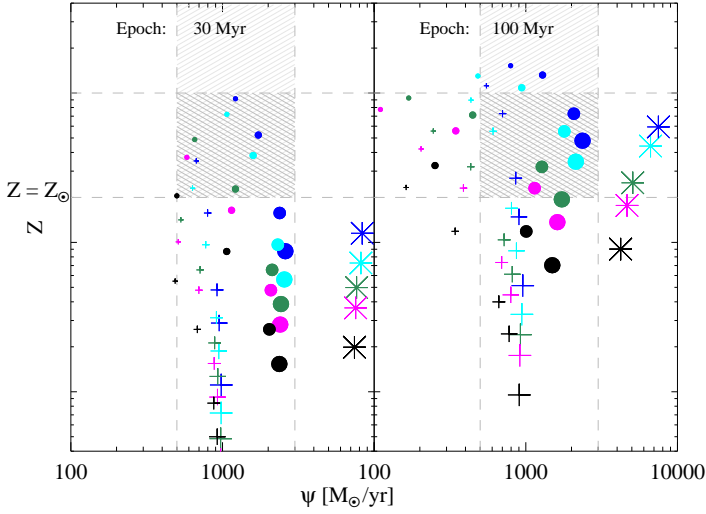}}       
      \caption{Relation between metallicity and SFR at epochs of 30 Myr (left panel) 
      and 100 Myr (right panel). 
      The colored symbols are obtained for different initial gas masses, $M_{\mathrm{ini}}$, 
      SFRs, and IMFs.
      The size of the symbols is scaled by $M_{\mathrm{ini}}$. 
      Calculations are made for   
       $M_{\mathrm{ini}}$ = 1.3 $\times$  $10^{12}$ $\Msun$ (largest symbol), 
       $M_{\mathrm{ini}}$ = 5 $\times$  $10^{11}$ $\Msun$,
       $M_{\mathrm{ini}}$ = 3 $\times$  $10^{11}$ $\Msun$,
       $M_{\mathrm{ini}}$ = 1 $\times$  $10^{11}$ $\Msun$, and
       $M_{\mathrm{ini}}$ = 5 $\times$  $10^{10}$ $\Msun$ (smallest symbol). 
      The crosses correspond to calculations for a initial SFR  
      $\psi_{\mathrm{ini}}$ =  $10^{3}$ $\Msun$ yr$^{-1}$, 
      the filled circles to $\psi_{\mathrm{ini}}$ = 3 $\times$ $10^{3}$ $\Msun$ yr$^{-1}$ 
      and the stars to $\psi_{\mathrm{ini}}$ = $10^{4}$ $\Msun$ yr$^{-1}$. 
      The black, green, cyan, magenta, and blue colors denote the Salpeter, 
      mass-heavy, top-heavy, Larson 1, and Larson 2 IMF, respectively. 
      The dark grey shaded region indicates the range of the metallicity and SFR 
      based on observations of QSOs at $z$ $>$ 6. 
      The vertical dashed lines represent the lower and upper limits 
      of the observationally derived SFRs.
      The light grey shaded area accounts for the uncertainty of the upper limit of the metallicity.
      The horizontal dashed lines mark the lower and possibly upper limit of the metallicity.
                              }
     \label{FIG:MZSFR}
   \end{figure}
We next present the obtained metallicities and SFRs at the time of observation for the models 
discussed above. 

Fig.~\ref{FIG:MZSFR} depicts the metallicity versus SFR at epochs of 30 Myr (left panel) 
and 100 Myr (right panel). 
With respect to observations of QSOs $>$ (5) 6 we marked the range of derived 
values as a dark grey shaded zone. 
The lower and upper limits of the SFR are based on observations by
 \citet{bertol03} and \citet{wan10}. 
We set the lower limit for the metallicity at the solar value and the upper limit at 5 $\Zsun$.
This is based on the inferred solar or supersolar metallicities in high-$z$ QSOs  
\citep[e.g.,][]{bar03, diet03, fan03, freu03, mai03, dimat04, bec06, juar09}. 
We note that there are no strong constraints on the upper limit and therefore the zone 
above 5 $\Zsun$ is marked as light grey shaded region to account for the uncertainty. 

We find that at an epoch of 30 Myr high metallicities in the less massive galaxies 
are already reached.  
The best result is attained by a system with 
$M_{\mathrm{ini}}$ = 1 $\times$  $10^{11}$ $\Msun$, 
$\psi_{\mathrm{ini}}$ = 3 $\times$ $10^{3}$ $\Msun$ yr$^{-1}$, 
and IMFs biased towards higher masses. 
For a galaxy with $M_{\mathrm{ini}}$ = 5 $\times$  $10^{10}$ $\Msun$ 
all models with either the same $\psi_{\mathrm{ini}}$ or with the lower initial SFR, and 
top-heavy IMFs are within the dark grey shaded region as well.

At an epoch of 100 Myr the metallicity has increased in all models, while the SFR 
in the less massive galaxies has significantly decreased. 
The models for $M_{\mathrm{ini}}$ = 3--5 $\times$  $10^{11}$ $\Msun$, 
$\psi_{\mathrm{ini}}$ = 3 $\times$ $10^{3}$ $\Msun$ yr$^{-1}$, and top heavy IMFs
 constitute the best results. 
In galaxies with  $M_{\mathrm{ini}}$ = 3 $\times$  $10^{11}$ $\Msun$, 
the same initial SFR, and either a mass-heavy or Larson 1 IMF the obtained 
values for $Z$ and  $\psi(t)$ are also in agreement with the observed values. 
The metallicities in the low mass galaxies which give the best agreement at 30 Myr 
are now shifted above the upper limit, while the SFRs remain in the observed range. 
The models for a galaxy with $M_{\mathrm{ini}}$ = 1 $\times$  $10^{11}$ $\Msun$, 
a lower initial SFR of $10^{3}$ $\Msun$ yr$^{-1}$, and top-heavy IMFs at this 
epoch (100 Myr) reach sufficiently high metallicities, while high enough SFRs are sustained.  
\subsection{CO conversion factor and gas-to-H$_2$ mass ratio}
\label{SEC:CONH2M}
%---------------------------------------------------------------------------------------------------------------------------------
%
%
    \begin{figure}
   \centering      
  \resizebox{\hsize}{!}{ \includegraphics{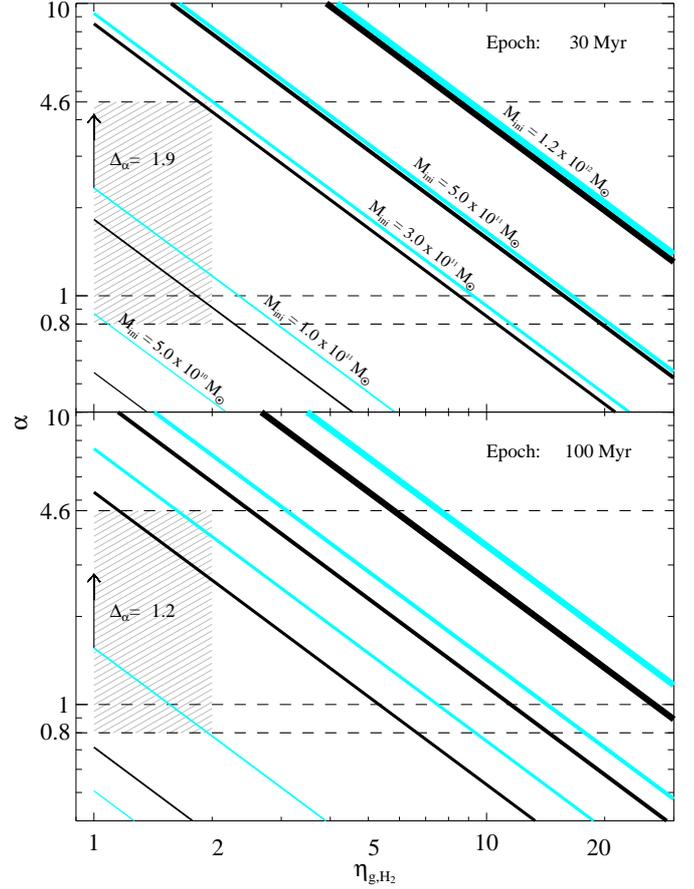}}      
      \caption{CO conversion factor versus gas-to-H$_2$ ratio at epochs 30 Myr (top panel) 
      and 100 Myr (bottom panel). 
      The solid lines correspond to calculations of $\alpha$ 
      as a function of the gas-to-H$_2$ ratio $\eta_{\mathrm{g,H_{\mathrm{2}}}}$
      for a CO  line luminosity of  
      $L'_{\mathrm{CO(1- 0)}}$ = 2.7 $\times$ 10$^{10}$ K km s$^{-1}$ pc$^{2}$. 
      Calculations are performed for different IMFs and galaxies for a range of different initial gas masses 
      $M_{\mathrm{ini}}$. 
      The thickness of the lines is scaled by $M_{\mathrm{ini}}$ as indicated in the upper panel.  
      The black and cyan colors denote the Salpeter and top-heavy IMF, respectively. 
      The arrow indicates the shift of $\alpha$ for calculations with the lower 
      $L'_{\mathrm{CO(1- 0)}}$ = 1.5 $\times$ 10$^{10}$ K km s$^{-1}$ pc$^{2}$, 
      and $\Delta_{\alpha}$ is the difference of $\alpha$ between the higher and lower $L'_{\mathrm{CO(1- 0)}}$.
      Calculations are shown for models with 
      $\psi_{\mathrm{ini}}$ = 3 $\times$ $10^{3}$ $\Msun$ yr$^{-1}$, 
      except for the model for the most massive galaxy for which $\psi_{\mathrm{ini}}$ = $10^{4}$ $\Msun$ yr$^{-1}$. 
      The grey shaded region signifies the possible range of $\alpha$ and 
      $\eta_{\mathrm{g,H_{\mathrm{2}}}}$. 
      The horizontal black dashed lines mark the values of $\alpha$ = 0.8, 1 and 4.6  $\Msun$ 
      (K km s$^{-1}$ pc$^{2}$)$^{-1}$. 
                      }
     \label{FIG:ALPHA}
   \end{figure}
To evaluate the calculated models, we additionally consider the relation between the gas-to-H$_2$ mass ratio and the CO conversion factor 
used to derive the molecular gas mass in a galaxy.

Detections of high excitation CO line emission in QSOs at $z$ $>$ (5) 6 indicate the 
presence of 0.7--2.5 $\times$ $10^{10}$ $\Msun$ of molecular hydrogen  
\citep[e.g.,][]{bertol03b, walt03, walt04, rie09, wan10}. 
This molecular gas mass is derived from the relation 
$M_{\mathrm{H_{\mathrm{2}}}}$ = $\alpha$ $\times$ $L'_{\mathrm{CO(1- 0)}}$, 
where $\alpha$ is the conversion factor between the low excitation 
CO J = 1--0 line luminosity $L'_{\mathrm{CO(1- 0)}}$ and  $M_{\mathrm{H_{\mathrm{2}}}}$. 
For spiral galaxies $\alpha$ is typically  
$\sim$ 4.6 $\Msun$ (K km s$^{-1}$ pc$^{2}$)$^{-1}$  \citep[e.g.,][]{sol91}, 
while for the centre of nearby ultra luminous starburst galaxies a conversion factor of 
$\alpha$ = 0.8--1 $\Msun$ (K km s$^{-1}$ pc$^{2}$)$^{-1}$ is appropriate
\citep[e.g.,][]{dow98}.  
The latter value of $\alpha$ is usually used for
e.g., 
high-$z$ QSOs \citep[e.g.,][]{bertol03b, walt03, wan10}, 
Ultra Luminous Infrared Galaxies (ULIRGs) \citep{yan10} 
or for high-$z$ sub-mm galaxies (SMGs) \citep{tec04, gre05}. 
However $\alpha$ is not well known  in the case of very high excitation. 

In our models we have computed the total (H + He) gas mass $M_{\mathrm{G}}$ 
which remains in the galaxies at a given epoch. 
The molecular gas mass, $M_{\mathrm{H_{\mathrm{2}}}}$, constitutes a certain fraction of the total gas mass, $M_{\mathrm{G}}$.
Hence we introduce the gas-to-H$_2$ mass ratio as 
$\eta_{\mathrm{g,H_{\mathrm{2}}}} = M_{\mathrm{G}} / M_{\mathrm{H_{\mathrm{2}}}}$. 
The CO conversion factor can thereby be expressed as a function of  
$\eta_{\mathrm{g,H_{\mathrm{2}}}}$ as
\begin{equation}
\label{EQ:METALL}
             \alpha =   \frac{ M_{\mathrm{G}} }{\eta_{\mathrm{g,H_{\mathrm{2}}}} \, L'_{\mathrm{CO(1- 0)}} } ,
\end{equation}
where $\eta_{\mathrm{g,H_{\mathrm{2}}}}$ $\ge$ 1 is kept as a free parameter.
In ULIRGs and SMGs a major fraction of the gas is believed to exist in form of 
molecular hydrogen \citep[e.g.,][]{sanmi96}. 
For example a value for $\eta_{\mathrm{g,H_{\mathrm{2}}}}$ of $\sim$ 1 has been found for the $z$ = 3
radio galaxy B3 J2330+3927 \citep{debreu03}. 
This might also be the case for QSOs and suggests a gas-to-H$_2$ ratio between 1 and 2.

In Fig.~\ref{FIG:ALPHA} we show the results for $\alpha$ as a function of  
$\eta_{\mathrm{g,H_{\mathrm{2}}}}$ with
$\psi_{\mathrm{ini}}$ = 3 $\times$ $10^{3}$ $\Msun$ yr$^{-1}$ for models with 
$M_{\mathrm{ini}}$ $\le$ 5 $\times$ $10^{11}$ $\Msun$ and with 
$\psi_{\mathrm{ini}}$ = $10^{4}$ $\Msun$ yr$^{-1}$ for the most massive galaxy. 
Calculations are performed for two different epochs; 30 Myr (top panel) and 100 Myr (bottom panel). 
The IMFs involved are the top-heavy IMF and the Salpeter IMF. 
We adopt a CO  line luminosity 
$L'_{\mathrm{CO(1- 0)}}$ = 2.7 $\times$ 10$^{10}$ K km s$^{-1}$ pc$^{2}$ 
which is based on the derived values of J1148+5251 and J0840+5624
 \citep[e.g.,][]{bertol03b, walt03, wan10}.
 
The difference of $\alpha$ from calculations with a lower $L'_{\mathrm{CO(1- 0)}}$ (i.e., $L'_{\mathrm{CO(1- 0)}}$ = 1.5 $\times$ 10$^{10}$ K km s$^{-1}$ pc$^{2}$) is indicated by the arrow in Fig.~\ref{FIG:ALPHA}.
The grey shaded area signifies a possible range for $\alpha$ and $\eta_{\mathrm{g,H_{\mathrm{2}}}}$ as discussed above.
 
For a fixed value of $\alpha$  the gas-to-H$_2$ ratio increases with increasing initial mass of the galaxy. 
This is as a consequence of the larger amounts of gas mass remaining in the more massive galaxies at the epochs of interest 
(see also  \citetalias{gall10a}). 
Conversely, for a fixed  $\eta_{\mathrm{g,H_{\mathrm{2}}}}$,  $\alpha$ increases with increasing $M_{\mathrm{ini}}$. 
The maximum value of $\alpha$ is obtained for  $\eta_{\mathrm{g,H_{\mathrm{2}}}}$ = 1, i.e., 
$M_{\mathrm{G}} \equiv M_{\mathrm{H_{\mathrm{2}}}}$. 
We find that at both epochs, the maximum value of $\alpha$ for the less massive galaxies is lower 
than $\sim$ 4.6 $\Msun$ (K km s$^{-1}$ pc$^{2}$)$^{-1}$. 
For a given $M_{\mathrm{ini}}$, $\alpha$, and  $\eta_{\mathrm{g,H_{\mathrm{2}}}}$  are lower at later epochs.
For a lower $L'_{\mathrm{CO(1- 0)}}$, 
$\alpha$ shifts to higher values for a given $\eta_{\mathrm{g,H_{\mathrm{2}}}}$.

At an epoch of 30 Myr the values for $\alpha$ and  $\eta_{\mathrm{g,H_{\mathrm{2}}}}$ are similar 
for all IMFs and galaxies with $M_{\mathrm{ini}}$ $>$ 1 $\times$  $10^{11}$ $\Msun$, 
while the difference becomes larger with decreasing $M_{\mathrm{ini}}$. 
Feasible values of $\alpha$ and $\eta_{\mathrm{g,H_{\mathrm{2}}}}$ are possible for galaxies with 
$M_{\mathrm{ini}}$ = 1 $\times$  $10^{11}$ $\Msun$ and the higher value of $L'_{\mathrm{CO(1- 0)}}$. 
For top-heavy IMFs  $\eta_{\mathrm{g,H_{\mathrm{2}}}}$ = 1 results in a maximum 
$\alpha$  of $\sim$ 2.3 $\Msun$ (K km s$^{-1}$ pc$^{2}$)$^{-1}$, 
while for $\alpha$ = 0.8 $\Msun$ (K km s$^{-1}$ pc$^{2}$)$^{-1}$, 
the fraction of molecular hydrogen is about one third of the total gas mass.
In the least massive galaxy ($M_{\mathrm{ini}}$ = 5 $\times$  $10^{10}$ $\Msun$) 
and for a top-heavy IMF  $\alpha$ $\approx$ 0.8 $\Msun$ (K km s$^{-1}$ pc$^{2}$)$^{-1}$ 
presupposes that all the gas in this system is in the form of molecular hydrogen. 
In more massive systems with $M_{\mathrm{ini}}$ = 1--3 $\times$ $10^{11}$ $\Msun$, a value of
$\alpha$ $\approx$ 0.8--1 $\Msun$ (K km s$^{-1}$ pc$^{2}$)$^{-1}$ presumes that the 
molecular hydrogen constitutes only a small  fraction of about 1/10--1/20 of the total gas mass.

At an epoch of 100 Myr a clear separation between the IMFs is noticeable. 
For a Salpeter IMF the galaxies underwent a stronger gas exhaustion than for a top-heavy IMF, 
which is more significant for the less massive galaxies.   
As for the epoch at 30 Myr the system with 
$M_{\mathrm{ini}}$ = 1 $\times$  $10^{11}$ $\Msun$ and top-heavy IMF is plausible
, i.e., for $\alpha$ $\sim$ 0.8 $\Msun$ (K km s$^{-1}$ pc$^{2}$)$^{-1}$ the gas-to-H$_2$ ratio 
$\eta_{\mathrm{g,H_{\mathrm{2}}}}$ = 2. 
For the galaxies with $M_{\mathrm{ini}}$ = 3--5 $\times$  $10^{11}$ $\Msun$ 
and top-heavy IMF we obtain $\alpha$ = 1.4--1.5 for a corresponding gas-to-H$_2$ ratio 
$\eta_{\mathrm{g,H_{\mathrm{2}}}}$ = 5--10, resulting in a molecular mass of 
$M_{\mathrm{H_{\mathrm{2}}}}$ $\sim$ 3.7 $\times$ $10^{10}$ $\Msun$.  
Alternatively, a higher value for  $\alpha$ up to 4.6 results in a lower $\eta_{\mathrm{g,H_{\mathrm{2}}}}$ = 2--4. 
It is noteworthy that for the assumed 
$L'_{\mathrm{CO(1- 0)}}$ = 2.7 $\times$ 10$^{10}$ K km s$^{-1}$ pc$^{2}$, 
$\alpha$ = 4.6 $\Msun$ (K km s$^{-1}$ pc$^{2}$)$^{-1}$ implies 
$M_{\mathrm{H_{\mathrm{2}}}}$ = 1.2 $\times$ $10^{11}$ $\Msun$. 
The likelihood that such a high $M_{\mathrm{H_{\mathrm{2}}}}$ could have been built up 
within a short timescale of 30--100 Myr however is unclear. 
%
%
%---------------------------------------------------------------------------------------------------------------------------------
% Section : Discussion
%---------------------------------------------------------------------------------------------------------------------------------
%   
\section{Discussion}
\label{SEC:DISC}
%---------------------------------------------------------------------------------------------------------------------------------
%
%
\begin{table*}%[bt]
\caption{{Observed properties of quasars at z $\gtrsim$ 6.0}}              
\label{TAB:OBS}
\centering
\begin{tabular}{lccccccl}
\hline
\hline
   Object		
   & $z$	
   &$L'_{\mathrm{CO(1- 0)}}$	
   & SFR	
   &$M_{\mathrm{d}}$	
   &$M_{\mathrm{H_{\mathrm{2}}}}$	
   &$M_{\mathrm{dyn}}\sin^{2}i$	
   &Ref. \\                 
   &		
   &10$^{10}$ K km s$^{-1}$ pc$^{2}$
   &$\Msun$ yr$^{-1}$
   &$10^{8}$ $\Msun$ 
   &$10^{10}$ $\Msun$ 
   &$10^{10}$ $\Msun$ 
   &\\
\hline

\\J1148+5251	& 6.42	& 3.0 $\pm$ 0.3	& 2380	& 5.9 $\pm$ 0.7	& 2.4 / 3.7\tablefootmark{a}  & 4.5	& 1,2,3,4\\
  J1048+4637	& 6.23	& 1.2 $\pm$ 0.2	& 650	& 4.3 $\pm$ 0.6	& 1.0 & 4.5	& 1,2,3\\
  J2054-0005	& 6.06	& 1.5 $\pm$ 0.3	& 1180	& 3.4 $\pm$ 0.8	& 1.2 & 4.2	& 5,2,3\\
  J0840+5624	& 5.85	& 3.2 $\pm$ 0.4	& 1460	& 4.7$\pm$ 0.9		& 2.5 & 24.2	& 6,2,3\\
\hline
\end{tabular}
\tablebib{
(1) \citet{fan03}; (2) \citet{wan10}; (3) \citet{mich10b}; 
(4)  \citet{walt04}; (5) \citet{jiang08}; (6) \citet{fan06}
}
\tablefoot{
\tablefoottext{a}{$M_{\mathrm{H_{\mathrm{2}}}}$ = 3.7 $\times$ $10^{10}$ $\Msun$ 
deduced from [\ion{C}{i}] line detections by \citet{rie09}}
}
\end{table*}
%\ \\
%
%
\begin{table*}%[bt]
\caption[Calculated properties from the best matching models of z $\gtrsim$ 6 QSOs from our sample]
{Calculated properties from the best matching models of z $\gtrsim$ 6 QSOs from our sample}        
\label{TAB:CALC}
\centering
\begin{tabular}{lcccclll}
\hline
\hline
         Object\tablefootmark{a}
     & SFR
     & $M_{\mathrm{d}}$
     & $M_{\mathrm{\ast}}$
     & $Z$
     & $\alpha$\tablefootmark{b}
     & $\eta_{\mathrm{g,H_{\mathrm{2}}}}$\tablefootmark{b}
     & $M_{\mathrm{H_{\mathrm{2}}}}$\tablefootmark{b} \\
     & $\Msun$ yr$^{-1}$
     &$10^{8}$ $\Msun$ 
     &$10^{10}$ $\Msun$
     &$\Zsun$
     &
     &
     &$10^{10}$ $\Msun$
 \\                 
\hline
\\ J1148+5251(A)	&1600	& 3.1--5.1  	& 3.5 	& 2	& 0.8--2.3 		& 3.0--1.0 		&2.16--6.2   \\
   J1148+5251(B)	&1000	& 2.4--8.9  	& 5.4		& 5	& 0.8--1.55	& 2.0--1.0		&2.10--4.1 \\
   J1048+4637(A)	&1000	& 2.4--8.9  	& 5.4		& 5	& 0.8--2.8 		&3.4--1.0 		&1.2--4.2  \\
   J1048+4637(B)	&610	& 3.5		  	& 2.8 	& 3.4	& 0.8--4.5		& 5.8--1.0 		& 1.2--6.7\\ 
   J2054-0005		&1150	& 2.7		  	& 4.7		& 4.4& 0.8--3.2 		& 3.0--1.0 		&1.2--4.8 \\
   J0840+5624(A)	&1500	& 2.1		  	& 11.0	& 4	& 0.8--4.6 		& 7.0--1.2 		&2.5--14.7   \\   
   J0840+5624(B)	&1400	& 4.8		  	& 20.0	& 5	& 0.8--4.6 		& 10--1.8 	  	&2.5--14.7  \\  
\hline
\\
\end{tabular}
\tablefoot{
\tablefoottext{a}{All models are calculated for a top-heavy IMF. Capital letters  in brackets (A,B) signify the different models (see corresponding models in Table~\ref{TAB:SUMA}) for the same object.}
\tablefoottext{b}{The ranges of $\eta_{\mathrm{g,H_{\mathrm{2}}}}$ and $M_{\mathrm{H_{\mathrm{2}}}}$ corresponds to the range of $\alpha$, which is between the commonly used value of $\alpha$ = 0.8 and the possible upper limit.}
}
\end{table*}
%\ \\
%
\subsection{Individual QSOs at  $z$ $\gtrsim$ 6}
\label{SEC:AIQ}
%---------------------------------------------------------------------------------------------------------------------------------
%
We ascertain plausible scenarios by comparing the model results discussed in 
Sect.~\ref{SEC:LOSB} with the derived values from observations for specific 
quantities of individual QSOs listed in Table~\ref{TAB:OBS}. 
The calculated values for diverse properties such as 
$M_{\mathrm{d}}$, $M_{\mathrm{\ast}}$, $M_{\mathrm{H_{\mathrm{2}}}}$, metallicity, and SFR from
the models discussed below, which best match the QSOs, are listed in Table~\ref{TAB:CALC}. 
The corresponding model parameters, and 
all models which match the discussed properties within the range defined by observations, 
are summarized in Table~\ref{TAB:SUMA}.

We find that at an epoch of 30 Myr the models with an initial mass of the galaxy of 
$M_{\mathrm{ini}}$ = 1 $\times$  $10^{11}$ $\Msun$, an initial SFR of 
$\psi_{\mathrm{ini}}$ = 3 $\times$ $10^{3}$ $\Msun$ yr$^{-1}$ and either a 
Larson 2 IMF, a top-heavy or a mass-heavy IMF 
reproduce the observed quantities of some QSOs at $z$ $>$ 6 in the case of a `maximum' SN efficiency.  

In particular, the model with a top-heavy IMF is best applicable to the QSO J1148+5251. 
The amount of dust reached is between 3.1--5.1 $\times$  $10^{8}$ $\Msun$  
for dust destruction in the ISM with $M_{\mathrm{cl}}$ = 100--0 $\Msun$. 
A stellar mass of $M_{\mathrm{\ast}}$ $\sim$ 3.5 $\times$  $10^{10}$ $\Msun$ is obtained.
The metallicity in the system is $\sim$ 2 $\Zsun$ and a SFR of $\sim$ 1600 $\Msun$ yr$^{-1}$ 
could be sustained. 
This model  is also favored given its values of $\alpha$ and $\eta_{\mathrm{g,H_{\mathrm{2}}}}$. 
The higher H$_2$ mass of $M_{\mathrm{H_{\mathrm{2}}}}$ = 3.7 $\times$ $10^{10}$ $\Msun$ 
derived by \citet{rie09} leads to  $\eta_{\mathrm{g,H_{\mathrm{2}}}}$ $<$ 2 and 
$\alpha$ $\sim$ 1.4 $\Msun$ (K km s$^{-1}$ pc$^{2}$)$^{-1}$.  
However, such a galaxy with $M_{\mathrm{ini}}$ = 1 $\times$  $10^{11}$ $\Msun$ 
implies that the dynamical mass is larger than the derived 
$M_{\mathrm{dyn}}$ of $\sim$ 5.5 $\times$  $10^{10}$ $\Msun$ (for a $i$ = 65$\degr$) by \citet{walt04}. 
While none of the models for $M_{\mathrm{ini}}$ = 5 $\times$  $10^{10}$ $\Msun$, 
which was used by  \citet{dwe07}, can be applied, a lower inclination angle similar to 
what has been adopted for the other QSOs might be considered.  

Another possible match with the properties of J1148+5251 is achieved by the 
same set of values for $M_{\mathrm{ini}}$, $\psi_{\mathrm{ini}}$, SN efficiency and IMF at an epoch of 100 Myr.
The calculated stellar mass is within the estimated range from observations and the dust mass 
is $\sim$ 2.4--8.9 $\times$  $10^{8}$ $\Msun$, depending on $M_{\mathrm{cl}}$. 
However, the SFR dropped to $\sim$ 1000 $\Msun$ yr$^{-1}$, while the metallicity increased to 
$\sim$ 5 $\Zsun$.  
In view of the lower SFR reached by these models  than suggested by observations at epochs either 30 or 100 Myr, 
a higher initial SFR than the 3 $\times$ $10^{3}$ $\Msun$ yr$^{-1}$ might be conceivable. 
In Fig.~\ref{FIG:MZSFR} one notices that a longer evolution with the same (or lower) initial 
SFR as used here does not lead to a better agreement with observations, since this results in 
an even lower SFR and higher metallicity. 

In view of this we find that this scenario at an epoch of 100 Myr is more appropriate for the 
QSOs J1048+4637  \citep{fan03} at $z$ = 6.23 and J2054-0005 \citep{jiang08} at $z$ = 6.06. 
For the latter QSO a fine tuning of the epoch to 70 Myr results in a better match. 
At this epoch we obtain a SFR of 1150 $\Msun$ yr$^{-1}$ and a metallicity of  $\sim$ 4.4 $\Zsun$. 
The amount of dust is 
$M_{\mathrm{d}}$ $\sim$ 2.7 $\times$  $10^{8}$ $\Msun$ (for $M_{\mathrm{cl}}$ = 100 $\Msun$), 
while the stellar mass is $M_{\mathrm{\ast}}$ $\sim$ 4.7 $\times$  $10^{10}$ $\Msun$. 
The lower derived $L'_{\mathrm{CO(1- 0)}}$  leads to 
$\eta_{\mathrm{g,H_{\mathrm{2}}}}$ $\sim$ 3--4 in case 
$\alpha$ = 0.8--1 $\Msun$ (K km s$^{-1}$ pc$^{2}$)$^{-1}$ is applied, 
while for $\eta_{\mathrm{g,H_{\mathrm{2}}}}$ $\sim$ 2 a value for  $\alpha$ of $\sim$ 1.6 
would be required. 
For J1048+4637 the model for a lower initial SFR of 
$\psi_{\mathrm{ini}}$ = $10^{3}$ $\Msun$ yr$^{-1}$ might be an option. 
The SFR is $\sim$ 610 $\Msun$ yr$^{-1}$ and the metallicity is $\sim$ 3.4 $\Zsun$. 
While the stellar mass remains low, $M_{\mathrm{\ast}}$ $\sim$ 2.8 $\times$  $10^{10}$ $\Msun$, 
a dust mass of $M_{\mathrm{d}}$ $\sim$ 3.5 $\times$  $10^{8}$ $\Msun$ 
is obtained for a `maximum' SN efficiency and moderate dust destruction in the ISM. 
However, for  $\alpha$ = 0.8--1 $\Msun$ (K km s$^{-1}$ pc$^{2}$)$^{-1}$ the gas-to-H$_2$ ratio is 
$\sim$ 5--6, since for the lower initial SFR the system at this epoch is less exhausted. 

At either the same or a later epoch the more massive galaxies with 
$M_{\mathrm{ini}}$ = 3--5 $\times$  $10^{11}$ $\Msun$, an initial SFR of 
$\psi_{\mathrm{ini}}$ = 3 $\times$ $10^{3}$ $\Msun$ yr$^{-1}$ and 
IMFs biased towards higher stellar masses are applicable to some $z$ $\sim$ 6 QSOs. 
The stellar mass, metallicity, and SFR of these systems are in agreement with observations, 
with either  top-heavy IMFs or a mass-heavy IMF leading to the best results.  
The amount of dust can be produced by SNe with a `high' SN efficiency and 
$M_{\mathrm{cl}}$ $\le$ 100 $\Msun$, although the dust masses reached are at the lower limit. 

At an epoch of 170 Myr the system with $M_{\mathrm{ini}}$ = 3 $\times$  $10^{11}$ $\Msun$ 
is plausible for the QSO J0840+5624 \citep{fan06} at $z$ = 5.85, if an inclination 
angle higher than the assumed 40$\degr$ is assumed.   
The SFR is $\sim$ 1500 $\Msun$ yr$^{-1}$ and the metallicity is $\sim$ 4 $\Zsun$. 
The stellar mass is around 1.1 $\times$  $10^{11}$ $\Msun$. 
The amount of dust obtained with a `high' SN efficiency is 2.1 $\times$  $10^{8}$ $\Msun$, 
while with the `maximum' SN efficiency the dust mass exceeds a few times  
$10^{9}$ $\Msun$ (as already at an epoch of 100 Myr).  
However, for a $L'_{\mathrm{CO(1- 0)}}$ = 3.2 $\times$ 10$^{10}$ K km s$^{-1}$ pc$^{2}$ 
as derived for this QSO
the gas-to-H$_2$ ratio of  $\eta_{\mathrm{g,H_{\mathrm{2}}}}$ $\sim$ 5--7 for 
$\alpha$ = 0.8--1 $\Msun$ (K km s$^{-1}$ pc$^{2}$)$^{-1}$ is higher than for the 
less massive galaxies. 
In case of a lower $\eta_{\mathrm{g,H_{\mathrm{2}}}}$  of $\sim$  2,  
$\alpha$  $\sim$  2.7 $\Msun$ (K km s$^{-1}$ pc$^{2}$)$^{-1}$ is required.  
The larger galaxy with $M_{\mathrm{ini}}$ = 5 $\times$  $10^{11}$ $\Msun$, 
$\psi_{\mathrm{ini}}$ = 3 $\times$ $10^{3}$ $\Msun$ yr$^{-1}$ and top heavy IMF 
can account for the observed quantities at an epoch of 400 Myr. 
The amount of dust reached with a `high' SN efficiency is $\sim$ 4.8 $\times$  $10^{8}$ $\Msun$ 
and the SFR is  $\sim$ 1400 $\Msun$ yr$^{-1}$. 
The metallicity and stellar mass are in agreement, but the fraction of  
$M_{\mathrm{H_{\mathrm{2}}}}$ is around 1/10 for  
$\alpha$ = 0.8 $\Msun$ (K km s$^{-1}$ pc$^{2}$)$^{-1}$, 
while  $\alpha$ $\sim$ 4 $\Msun$ (K km s$^{-1}$ pc$^{2}$)$^{-1}$ is needed for 
$\eta_{\mathrm{g,H_{\mathrm{2}}}}$ of $\sim$  2. 
A higher amount of $M_{\mathrm{H_{\mathrm{2}}}}$ as denoted by the higher value of  $\alpha$ in these massive galaxies might be possible. For example, the presence of large amounts of cold and low-excited molecular gas have been suggested 
by \citet{pap01} for the QSO APM 08279+5255 at $z$ = 3.91. 
\begin{table*}[t!]
\caption[Models which match the observed range of properties of $z$ $\ge$ 5 QSOs]
{Models\tablefootmark{a} which match the observed range of properties of $z$ $\ge$ 5 QSOs}.         
\label{TAB:SUMA}
\centering
\begin{tabular}{llcclll}
\hline
\hline
    Epoch
    &$M_{\mathrm{ini}}$
    &$\psi_{\mathrm{ini}}$
   & SN efficiency
   & $M_{\mathrm{cl}}$
    & IMF 
    &z $\gtrsim$ 6 QSOs from our sample\tablefootmark{b} \\
     $\Msun$
   & 10$^{3}$ $\Msun$ yr$^{-1}$
   &
   &$\Msun$
   & 
 \\                 
\hline\\
\hline
 30 Myr&
 5 $\times$ $10^{10}$ 	&3 	&max	&100		&top-heavy, Larson 1, 2, mass-heavy&\\
				 	&&3 	&max	&0			&top-heavy, Larson 1, 2&\\
					&&{\bf3} 	&{\bf max}	& {\bf 0}			&{\bf mass-heavy}&\\
					&&1	&max	&0,100		&top-heavy, Larson 2&\\
 
  &1 $\times$ $10^{11}$ 	&{\bf3 }	&{\bf max}	& {\bf 100}	&{\bf Larson 2, mass-heavy} &\\
  					&&{\bf3 }	&{\bf max}	&{\bf 0}		&{\bf mass-heavy}&\\
					&&{\bf3 }	&{\bf max}	& {\bf 100--0}	&{\bf top-heavy} &J1148+5251(A)\\
					&&3 	&high	&0		&Larson 2&\\
70 Myr&
  1 $\times$ $10^{11}$ 	&{\bf1 }	&{\bf max}	& {\bf 100--0}	&{\bf top-heavy} &J2054-0005\\
100 Myr&
 5 $\times$ $10^{10}$ 	&3 		&max	&100	& Larson 2&\\
					&&{\bf3} 	&{\bf max}	&{\bf 0}	&{\bf top-heavy,  Larson 2}&\\
					&&1 		&max	&100	& Larson 2&\\

  &1 $\times$ $10^{11}$ 	&{\bf3 }	&{\bf max}	&{\bf 100}	&{\bf Larson 2}&\\
  					&&3 		&high	&0		&  Larson 2&\\
					&&{\bf3 } 	&{\bf max}	&{\bf 100--0}	&{\bf top-heavy}&J1148+5251(B), J1048+4637(A) \\
					&&1 		&max	&100	&Larson 2&\\
					&&{\bf1 }	&{\bf max}	&{\bf 100--0}	&{\bf top-heavy}&J1048+4637(B) \\

  &3 $\times$ $10^{11}$ 	&3 		&max	&800	& Larson 1, mass-heavy&\\ 	 	
					&&{\bf3} 	&{\bf max}	&{\bf800}	&{\bf top-heavy, Larson 2}&\\ 
					&&3 		&max	&100	&Larson 1, mass-heavy&\\ 										
  					&&3 		&high	&0		& mass-heavy&\\ 
					&&{\bf3} 	&{\bf high}	&{\bf0}	&{\bf top-heavy, Larson 2}&\\
					&&{\bf1} 	&{\bf max}	&{\bf800}	&{\bf Larson 2}&\\
					&&1 		&high	&0		& Larson 2&\\
  
  &5 $\times$ $10^{11}$ 	&3 		&max	&800	& mass-heavy&\\ 
					&&{\bf3} 	&{\bf max}	&{\bf800}	&{\bf top-heavy, Larson 2}&\\ 
					&&3 		&high	&100	&top-heavy, mass-heavy&\\
					&&{\bf3} 	&{\bf high}	&{\bf100}	&{\bf  Larson 2}&\\
					&&3 		&high	&0		& mass-heavy&\\
					&&{\bf3} 	&{\bf high}	&{\bf0}	&{\bf top-heavy, Larson 2}&\\

170 Myr&
  3 $\times$ $10^{11}$ 	&{\bf1 }	&{\bf high}	& {\bf 100--0}	&{\bf top-heavy} &J0840+5624(A) \\

400 Myr&
  5 $\times$ $10^{11}$ 	&{\bf1 }	&{\bf high}	& {\bf 100--0}	&{\bf top-heavy} &J0840+5624(B)  \\
\hline
\\
\end{tabular}
\tablefoot{
\tablefoottext{a}{All models which match the observed range of all properties of $z$ $\ge$ 6 QSOs are marked in bold face.}
\tablefoottext{b}{Capital letters in brackets signify that different models (A,B) are applicable for the same object}
}
\end{table*}
%\ \\
%
\subsection{SN efficiency and mass of the galaxy}
\label{SEC:MVH}
%---------------------------------------------------------------------------------------------------------------------------------
%
Our calculations show that with increasing $M_{\mathrm{ini}}$ (and fixed $\psi_{\mathrm{ini}}$, IMF) the SN dust production efficiencies can either be lowered or the degree of dust destruction increased in order to reach the required large dust masses.
This is best demonstrated by models for the most massive galaxies with 
$M_{\mathrm{ini}}$ = 3--13 $\times$  $10^{11}$ $\Msun$ in which  a `high' SN efficiency is sufficient in case of moderate to no dust destruction.

However, the largest system with $M_{\mathrm{ini}}$ = 1.3 $\times$  $10^{12}$ $\Msun$
exceeds the plausible dynamical masses derived from 
observations of QSOs at $z$ $\gtrsim$ (5) 6 by more than an order of magnitude.  
Moreover, our computed models show that at least one of the properties of either 
SFR, $Z$ or $M_{\mathrm{\ast}}$ are not in agreement with observations at any 
epoch for any assumption of either the initial SFR or the IMF  \citepalias[see also][]{gall10a}.  
Additionally the values for $\eta_{\mathrm{g,H_{\mathrm{2}}}}$ remain very high even for 
$\alpha$ = 4.6 $\Msun$ (K km s$^{-1}$ pc$^{2}$)$^{-1}$.
We therefore conclude that such a massive system as  
advocated by  \citet{val09}, cannot be applied to QSOs at $z$ $>$ (5) 6. 
Although systems with $M_{\mathrm{ini}}$ = 3--5 $\times$ $10^{11}$ $\Msun$ are appropriate for some QSOs at $z$ $<$ 6, 
such massive systems can only be applied to QSOs $>$ 6 when the inclination angle is lower than the assumed average angle. 

The models which best reproduce the observed properties of QSOs $>$ 6 are for a galaxy with $M_{\mathrm{ini}}$ = 1 $\times$ $10^{11}$ $\Msun$, but necessitate a `maximum' SN efficiency and/or a moderate amount of dust destruction. 
The overall rapid evolution of dust and some properties in these models indicates that such QSOs could possibly be present at a higher redshift than $z$ $>$ 6.4.  
An interesting example at a lower redshift of $z$ = 1.135 is the ULIRG SST J1604+4304, 
which shows properties similar to the considered high-$z$ QSOs. 
\citet{kawa10} reported a dust mass in this ULIRG of 1--2 $\times$ $10^{8}$ $\Msun$, 
a metallicity of around 2.5 $\Zsun$ and estimated the age of the stellar population to be 40--200 Myr. 

The possibility of moderate dust destruction in the ISM was already discussed in \citetalias{gall10a}. 
We found that the amount of dust for most models better coincide with observations for 
$M_{\mathrm{cl}}$ $\le$ 100 $\Msun$, which would be in agreement with the values of 
$M_{\mathrm{cl}}$ of 50--70 $\Msun$  derived for a multiphase ISM  \citep[e.g.,][]{mct89, dwe07}.  

The `maximum' SN efficiency might be problematic. 
There is only little observational evidence that SN can be very efficient  
\citep[e.g.,][]{wils05, dou01b, dun09}, 
and theoretical models predict  significant dust destruction 
in reverse shocks of SNe \citep[e.g.,][]{bia07, noz07, noz10}. 
On the other hand, these models also show that the effectiveness of dust destruction depends 
on various properties such as the geometry of the shocks, the density of the ejecta and the ISM, 
the size and shape of the grains, clumping in the SNe ejecta, and different SN types.   
In addition there is some  observational evidence that Type IIn SNe and sources such as 
luminous blue variables are possibly efficient dust producers \citep{fox09, smi09, gom10}. 
While dust production and destruction in SNe is yet unresolved, 
a `maximum' SN efficiency cannot be ruled out 
(e.g., Gall et al. in prep). 
Alternatively, either dust formation in the outflowing winds of QSOs  or 
grain growth in the ISM might be an option \citep[e.g.,][]{elv02, dwe07, drai09, mich10b, pip10, dwe10}
as supplementary or primary dust sources. 
However it remains to be investigated, if dust grain growth can be as efficient as required 
under the prevailing conditions of high star formation activity and a short time span. 
Typical grain growth timescales in molecular clouds are of order $10^{7}$ yr, but depending on the
density and metallicity these can possibly be shorter \citep[e.g.,][]{hiras00, zhuk08, drai09}. 
The fact that the starburst is assumed to occur in an initially dust free galaxy implies that 
heavy elements first need to be ejected into the ISM before grain growth can take place. 
In forthcoming work we will further develop the model to investigate the 
impact of different infall and outflow scenarios on the evolution of the 
amount of dust and various properties of a galaxy.    
%
%---------------------------------------------------------------------------------------------------------------------------------
% Section 6: Acknowledgements 
%---------------------------------------------------------------------------------------------------------------------------------
%
 \begin{acknowledgements}
 
We would like to thank Michal Micha{\l}owski, Darach Watson, Thomas Greve, and Sabine K\"onig 
for informative and helpful discussions. 
We also thank the anonymous referee for useful suggestions which helped improve the paper.
The Dark Cosmology Centre is funded by the DNRF.
            
\end{acknowledgements}
%
%---------------------------------------------------------------------------------------------------------------------------------
% Section 6: Reference 
%---------------------------------------------------------------------------------------------------------------------------------
%
%
\bibliographystyle{aa}
\bibliography{reflist_ch}
%
%---------------------------------------------------------------------------------------------------------------------------------
% End with document
%---------------------------------------------------------------------------------------------------------------------------------
%
\end{document}